\documentclass{aa}
\usepackage{graphicx}
\usepackage{txfonts}
\usepackage{natbib}

\bibpunct{(}{)}{;}{a}{}{,}

\begin{document}


   \title{The Baryonic Tully-Fisher relation revisited}

   \author{D. Pfenniger \and Y. Revaz}

   \offprints{D. Pfenniger}
   
   \institute{Geneva Observatory, University of Geneva, CH-1290
     Sauverny, Switzerland}

   \date{Received -- -- 20--/ Accepted -- -- 20--}
   
   \abstract{ The Baryonic Tully-Fisher relation (BTF) can be
     substantially improved when considering that the galactic
     baryonic mass is likely to consist not only of the detected
     baryons, stars and gas, but also of a dark baryonic component
     proportional to the HI gas.  The BTF relation is optimally
     improved when the HI mass is multiplied by a factor of about 3,
     but larger factors up to $11-16$ still improve the fit over the
     original one using only the detected baryons.  The strength of
     this improved relation is quantified with up-to-date statistical
     tests such as the Akaike Information Criterion or the Bayesian
     Information Criterion.  In particular they allow us to show that
     supposing a variable $M_\star/L$ ratio instead is much less
     significant. This result reinforces the suggestion made in
     several recent works that mass within galactic disks must be a
     multiple of the HI mass, and that galactic disks are
     substantially, but not necessarily fully, self-gravitating.
      
     \keywords{kinematics and dynamics of galaxies -- Tully-Fisher
       relation -- dark matter } 
    }

   \maketitle

\section{Introduction}

\citet{mcgaugh00} (hereafter MSBD) have extended the Tully-Fisher
relation \citep{tully77} (TF) over 5 dex in stellar mass and 1 dex in
velocity by correlating not the luminosity with the rotation velocity,
but the {\it detected\/} baryon mass with the rotation velocity.  This
is an important step toward understanding the TF relation, because the
physical causality link between mass and rotational velocity,
gravitational dynamics, is much more direct than between stellar light
production and rotational velocity.  The traditional TF relation
requires us to understand how the nuclear energy production inside stars
may well be tightly correlated with the global rotation speed of the
galaxy, while the baryonic TF (BTF) just requires finding the link
between the baryonic mass and kinematics.

The baryons considered by MSBD include the stars, the total mass of
which is inferred from a plausible stellar mass to light ratio, and
neutral hydrogen HI from 21\,cm measurements, including a cosmic
helium fraction.  Doing so, MSBD show with a sample of 243 galaxies
that a scatter plot of velocity--mass looks straighter when summing the
HI mass to the stellar mass.  The argument advanced by MSBD is that a
better relation exists between these baryons and rotational velocity
than between light and rotational velocity because the low surface
brightness galaxies in the sample have a very low stellar mass
content.  This makes sense because in the limit of a pure gaseous
protogalactic disk no stars shine, yet the disk does rotate.

Several reasons motivated us to reexamine this work.  By naming the
improved relation ``baryonic'' MSBD implies that the baryons are well
represented by the detected stars and HI, while in many spirals often
as much gas as the detected HI but in the form of H$_2$ can be
inferred from CO observations with the conventional proportionality
$X$ factor between the CO emission and the H$_2$ mass
\citep{combes04}.  This point is however not necessarily crucial
because the CO emission is often roughly located in the optical disk,
so the CO related H$_2$ mass may be well absorbed, to a first approximation,
in the mass to light ratio of the stellar mass.  But for
years now the universality of the $X$ factor has been challenged.
Recent rediscussions of this factor \citep[e.g.,][]{boulanger04} favor
an increased $X$ in low metallicity, low excitation or cold
conditions, typical of outer galactic HI disks, by at least one order
of magnitude.

Long ago \citet{bosma81a} pointed out that the dynamical mass in
spirals is known to be well correlated to a multiple of the detected
HI \citep[see also][]{hoekstra01}. So a natural hypothesis reinforced
by a series of arguments presented in \citet{pfenniger94a}, and
\citet{pfenniger94b} as well as variants discussed by other authors
\citep{henriksen95,depaolis95,gerhard96} is to assume that a
sufficiently cold form of H$_2$ can well make up a sizable fraction of
the unseen baryonic mass in the disk. Moreover, several new
theoretical and observational arguments indicate that galactic disks
may be more massive than usually thought. \citet{kalberla03} found
that the mass distribution and the dynamics of the Milky Way is
dominated by a dark matter disk. Massive disks are also needed to
explain spiral structures in low surface brightness galaxies
\citep{fuchs03,mayer04}, as well as in extended HI disks like in the
blue compact dwarf galaxy NGC\,2915 \citep{masset03}.  In the context
of warped galaxies, \citet{revaz04} have recently shown that the high
number of warped spirals results naturally from bending instabilities,
if the disk contributes to at least $60$\% of the total galactic mass,
within $30\,\rm{kpc}$.

The true number of baryons contained in spiral
galaxies can still be a multiple of the detected ones.  Presently the
hypothesis that MACHOS can represent an important part of the unseen
baryonic mass in an extended halo seems to be disproved by the
micro-lensing experiments towards the Magellanic clouds and the
Galactic center \citep{alcock01,afonso03}, but a dark baryonic
component closer to the outer disk, i.e., more closely related to the
HI distribution, has not been investigated to a similar extent yet.

If some sizable fraction of the mass correlates well with velocity,
and the detected baryons within the radius of measured velocities do
represent a non-negligible amount of the total mass within this
radius, it seems reasonable to check whether the full dynamical mass
could not improve the correlation.  Indeed, gravitational dynamics is
little dependent on the nature of matter over a few dynamical time-scales.

In order to investigate better the underlying apparent correlation,
the analysis of the MSBD sample can easily be extended by up-to-date
statistical tools.  Below, we use statistical tools, mainly the Akaike
Information Criterion \citep{akaike74}, and the Bayesian Information
Criterion \citep{schwarz78} that are not well known in astronomy, but
have been often used in many other scientific fields over the last two
decades.  They provide interesting decision criterions for
discriminating quantitatively between models using a different number
of fitting parameters for the same data set.  The curious reader will
find in \cite{liddle04} a short introduction for astronomers to these
methods, and an application to the cosmological parameters.

The purpose of the following Sections is therefore to extend the
discussion about the BTF with more advanced statistical techniques
than the ones currently popular in astronomy.


\section{Sample}\label{sample}


MSBD kindly provided us the sample that they used in their work, and
to make a reasonable comparison we use here exactly the same sample,
including the same assumptions about the mass-to-light ratios of the
stellar and HI components%
\footnote{In the following, and as commonly assumed in HI related
  works, by HI mass one means the 21\,cm optically thin
  measured HI mass augmented by the cosmic He abundance.}.  %
The sample consists of 243 approximately face-on spirals with peak
rotational velocities in the range 33 to $323\,\rm km\,s^{-1}$, almost
a dex.  For a more detailed description of the sample we refer the
reader to the MSBD paper.


\section{Optimized HI scaling}\label{sec2}


\subsection{Full sample}\label{full_sample}

\begin{figure}
  \resizebox{\hsize}{!}{\includegraphics[]{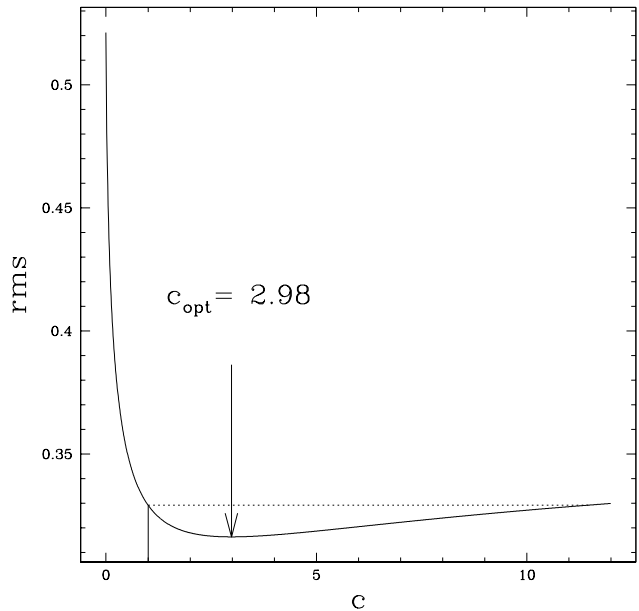}}
  \caption{Rms of the linear least-squares fit as a function of the 
    parameter $c$ (see Eq.~\ref{eq1}). The dotted line indicates
    MSBD's rms value ($c=1$).}
  \label{opt0}
\end{figure}

\begin{figure}
  \resizebox{\hsize}{!}{\includegraphics[]{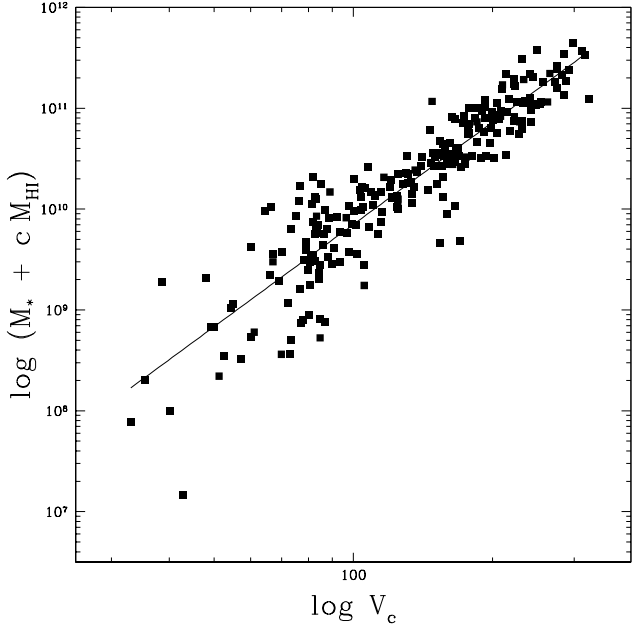}}
  \caption{Least-squares fit of the Baryonic Tully-Fisher relation using
    Eq.~\ref{eq1} with $a=3.11$, $b=3.36$ and $c=2.98$.}
  \label{btfi}
\end{figure}

The first question is how scaling HI with a constant factor $c$
different from $1$ changes the fit rms.  Using a linear least-squares
fit, for $c$ in the range $[0-15]$, we calculate the corresponding
coefficients $a$ and $b$ in the least-squares relation linking the
stellar mass $M_\star$ inferred from the light and a constant
$M_\star/L$ ratio, the HI mass $M_{\rm HI}$, and the rotational
velocity $V_{\rm c}$:
\begin{equation}
  \log (M_\star + c M_{\rm HI}) \approx a + b \log V_{\rm{c}}\, , 
  \label{eq1}
\end{equation} 
as well as the rms of the respective fits as a function of $c$, which
are plotted in Fig.~\ref{opt0}.  Clearly, passing from $c=0$
($\rm{rms}=0.521$) in the usual TF relation to $c=1$
($\rm{rms}=0.329$) in the BTF relation decreases the fit rms by a
factor of $1.6$, which is the main new point in MSBD's paper.  Yet the
curve rms($c$) continues to decrease for $c>1$ down to a shallow
minimum at $c=2.98$ ($\rm{rms}=0.3164$).  Beyond this optimal $c$, the
rms grows slowly, such that up to $c=11.5$ the rms fit is still better
than for $c=1$.
Fig.~\ref{btfi} shows the least-squares fit of the extended BTF
(hereafter EBTF) relation at the optimal value $c_{\rm opt}=2.98$, where the
linear fit parameters are $a=3.11$ and $b=3.36$.
%

\subsection{Comparing models}\label{comparing}

With an additional parameter, we naturally expect that the rms of the
EBTF model (Eq.~\ref{eq1}) decreases with respect to the simpler BTF
model (null hypothesis), where $c$ remains fixed to $1$. Thus, even if
the rms is lower, it does not mean that the data are better fitted by
the EBTF model.

To test this, we have used three different statistical tools.  The
first tool is classical, while the two more recent ones are less well
known in astronomy, but have become popular elsewhere.  The methods work
under the assumption that the error distribution is Gaussian.  A study
of the rms distribution shows that the data points do verify
reasonably well this assumption.

\subsubsection{Extra sum-of-squares $F$-Test}

This standard test consists of comparing the difference between the
$\chi2$ sums of two models, say model 1 and model 2, where model 1 is
a \emph{simpler case} of model 2 %
\footnote{By \emph{simpler case}, we mean that model 1 is identical to
  model 2, with a number $r$ of parameters fixed. The two models are
  also said to be \emph{nested}.  Since in MSBD's BTF model the
  parameter $c$ is fixed to $1$, this model is a \emph{simpler model} 
  than the EBTF one.}. %
Because model 1 has a lower number of parameters, we usually expect
that $\chi2_1$ is greater than $\chi2_2$ with a difference
corresponding to the difference $r$ of parameters between the two
models. Thus, the ratio $\left( \chi2_1-\chi2_2 \right)/r$ provides
an estimate of the goodness of one model compared to the other.  An
$F$-value is defined by dividing the previous ratio by the normalised
fraction $\chi2_2/(N-k_2)$, where $N-k_2$ is the degrees of freedom
of model 2 with $k_2$, its number of parameters. Explicitly, $F$ is
written as:
\begin{equation}
  F = 
  \frac{ \left( \chi2_1-\chi2_2 \right) }{r} \,\Bigg/\, 
  \frac{\chi2_2}{(N-k_2)} \, .
        \label{f}
\end{equation} 
Since $F$ is a ratio of two $\chi2$ distributions, it follows an
$F$-distribution.  We can then compute the probability $P$ to find a
value greater than $F$ for such a distribution. If this probability is
weak (typically less than 5\%), we can conclude that the $F$-value
obtained has a weak probability to result from the error fluctuations
and that the data is significantly better fitted by model 2.

Assuming that the intrinsic errors $\sigma_i$ are constant for each
point $i$ ($\sigma_i=\sigma$), the $\chi2$ terms in Eq.~(\ref{f}) can
be replaced by the sum-of-squares $S2_i$:
\begin{equation}
  {S2_i} = \sum_{j=1}^{N}\left( y_j - f_i(x_j) \right)2 \, ,
 \label{ss}
\end{equation} 
where the $y_j$ correspond to the left hand side of Eq.~(\ref{eq1})
and the $f_i(x_j)$ to the right hand one. A rigorous description of
this classical test can be found, e.g., in
\citet[][p.~98-101]{lupton95}.

Comparing the EBTF model with the simpler BTF model, with $N=243$,
$k_1=2$ and $r=1$, we obtain $F=19.9$, to which the associated
probability $P$ is lower than $0.01$\%, which is the probability for
the BTF model to be better than the EBTF one.

\subsubsection{Akaike's Information Criterion corrected (AICc)}

The second performed test is derived form the Information theory,
namely the Akaike's Information Criterion corrected (AICc)
\citep{sakamoto86,burnham04}.  Since the formal derivation of this
test is somewhat lengthy we refer the reader to the literature for
more details \citep[see, e.g.,][]{brockwell02,burnham04}.  However,
its application is as simple as the Extra sum-of-squares $F$-Test,
without the need to introduce additional assumptions%
\footnote{On the contrary, the AICc test is less restrictive because
  it does not require \emph{nested} models.}.  %
The AICc consists of computing for the two models the values:
\begin{equation}
 \mathcal{A}^{\rm c}_i = 
     N\,\ln \left( \frac{S2_i}{N}\right) + 2\,k_i +  
                 \frac{2\,k_i\left(k_i+1\right)}{N-k_i-1} \, ,
\end{equation} 
where $k_i$ is the number of parameters of model $i$ and $S2_i$ is
the corresponding sum-of-squares defined by Eq.~(\ref{ss}).  In this
formula, the first term is the entropy term, which usually decreases
when the number of parameters $k_i$ increases.  It is balanced by the
second term linear in $k_i$.  The third term is the correction term
especially relevant when $N$ is small, so unimportant here.  As for
entropy in statistical mechanics, the value of $\mathcal{A}^{\rm c}_i$
depends on the choice of data unit, therefore its value has no
absolute meaning, only the difference between two models has one: the
relative information content.  The best model is simply the one with
the lowest $\mathcal{A}^{\rm c}_i$ value, where the decrease of the
entropy term dominates.  The difference $\Delta \mathcal{A}^{\rm{c}} =
\mathcal{A}^{\rm{c}}_2-\mathcal{A}^{\rm{c}}_1$ allows us to quantify
how much a model is better than another.  The probability
$\mathcal{P}_{2,1}$ that model 2 is better than model 1 is given by:
\begin{equation}
 \mathcal{P}_{2,1} = 
    \frac{ e^{-\frac{1}{2}\,\Delta \mathcal{A}^{\rm{c}}} }
         { 1 + e^{-\frac{1}{2}\,\Delta \mathcal{A}^{\rm{c}}} }.
 \label{eq3}
\end{equation} 

The comparison of the BTF model with the EBTF one gives respectively
$\mathcal{A}^{\rm{c}}_1 = -533.79$ and $\mathcal{A}^{\rm{c}}_2 =
-551.14$.  The so called ``information ratio''
$\mathcal{P}_{2,1}/\mathcal{P}_{1,2}$ tells us that the EBTF model is
5854 times more likely to be correct than the BTF one.

\subsubsection{Bayesian Information Criterion (BIC)}

Another criterion which is often less favorable for models with many
parameters is the Bayesian Information Criterion (BIC)
\citep{schwarz78}.  This criterion, favoured by \cite{liddle04},
weights more heavily additional model parameters than AICc for data
sets with large $N$.  As AICc, BIC does not require us to compare nested
models.  The BIC consists of computing for two models the values:
\begin{equation}
 \mathcal{B}_i = 
     N\,\ln \left( \frac{S2_i}{N}\right) + k_i\ln N \, ,
\end{equation} 
where the parameters have the same meaning as before.  As for AICc,
only the difference of $\mathcal{B}_i$ is important, not their
particular value.

The comparison of the BTF model with the EBTF one gives respectively
$\mathcal{B}_1=-523.41$ and $\mathcal{B}_2=-537.34$.  Applying the
analogue of Eq.~(\ref{eq3}) for BIC yields that the EBTF model is 1054
times more likely to be correct than the BTF one.

\subsubsection{Comparison with a quadratic fit}

In order to convince ourself of the strength of the three previous
comparison methods, we have compared the BTF model with a third model,
where the data is simply fitted with a quadratic function, therefore
with the same number of free parameters ($k=3$) as our model:
\begin{equation}
   \log (M_\star + M_{\rm HI}) \approx
      a + b \log V_{\rm{c}} + c^\prime \log2 V_{\rm{c}} \, .
   \label{quadratic}   
\end{equation} 
Comparing this quadratic model with the BTF model, the $F$-test gives
a probability $P$ of only $37.9$\%.  In this case, the weak decrease
of the rms of the more complex model is not sufficient to justify the
additional parameter $c^\prime$.  The AICc $\mathcal{A}^{\rm c}_i$ for
the two models are respectively $-533.79$ and $-532.51$; the BIC
$\mathcal{B}_i$, $-523.41$ and $-518.70$ respectively: the BTF model
is clearly a slightly better model than the quadratic fit.

Since a quadratic model does not significantly improve the fit, we can
then conclude that the improvement reached by the EBTF model does not
result merely from the additional parameters but rather from the
judicious choice of the fitting law Eq.~(\ref{eq1}), directly
inspired by physical considerations.

\subsection{Subsamples} 

In this Section, we check the robustness of the result obtained in
Sect.~\ref{sec2} when removing points from the the MSBD data.  All the
subsamples containing all but one or two points are analysed.

\begin{figure}
        \resizebox{\hsize}{!}{\includegraphics[]{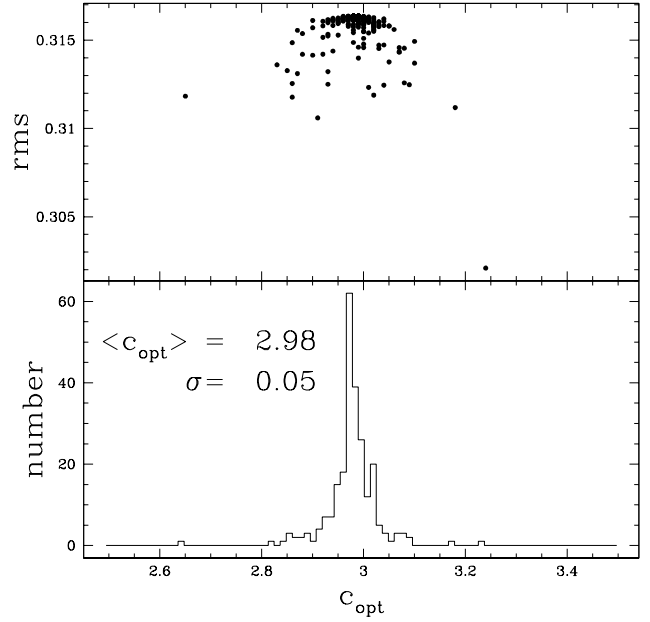}}
        \caption{
          \textbf{Bottom}: Histogram of the optimal $c_{\rm opt}$
          values obtained by fitting Eq.~(\ref{eq1}) to successively
          the 243 distinct subsamples made of 242 data points.
          \textbf{Top}: Rms of each subsample fit.  }
        \label{opt1}
\end{figure}

For each of the $N=243$ subsamples, where in each one a single point
has been successively removed, we have computed the optimal
$c_{\rm{opt}}$ value, as described in Sect.~\ref{full_sample}. The
histogram of those values is displayed at the bottom of
Fig.~\ref{opt1}.  The values $c_{\rm{opt}}$ lie in the interval $2.65$
and $3.24$ with a mean of $2.98$ and a dispersion of $0.05$.
Therefore the optimal $c_{\rm{opt}}$ is little sensitive to the error
produced by any single data point.

\begin{figure}
   \resizebox{\hsize}{!}{\includegraphics[]{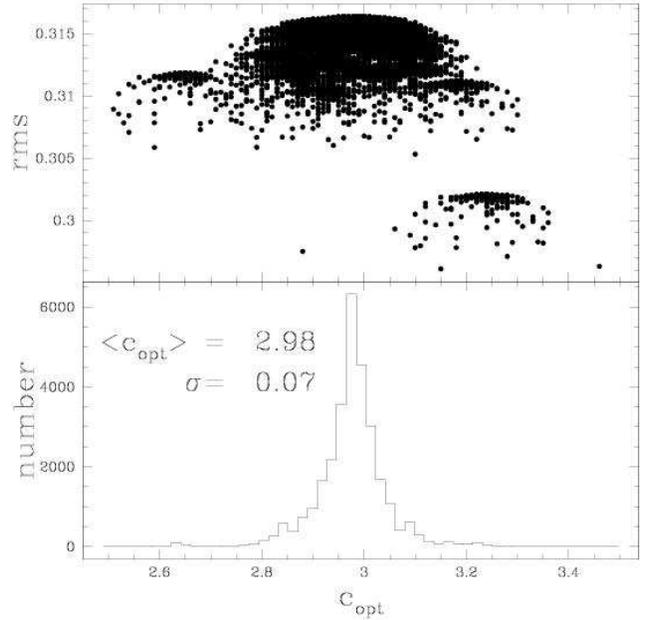}}
   \caption{
     \textbf{Bottom}: Histogram of the optimal $c_{\rm{opt}}$ values
     obtained by fitting Eq.~(\ref{eq1}) to each of the distinct
     29403 subsamples made of 241 data points.
     \textbf{Top}: Rms of each subsample fit.  }
        \label{opt2}
\end{figure}

The same exercise is repeated by successively removing two different
points for each distinct combination allowed by 243 data points.
Thus, $243\times 242/2=29403$ searches for an optimal value of
$c_{\rm{opt}}$ have been computed.  The found $c_{\rm{opt}}$ range
between $2.51$ and $3.46$ with a mean of $2.98$ and a dispersion of
$0.07$.

In the rms of both histogram peaks, we observe that a particular
removed point contributes to a particularly large decrease of the rms
and an increase of $c_{\rm opt}$.  This point corresponds to the
isolated lower left point of Fig.~\ref{btfi}. It is important to
emphasize that the conclusions obtained in the previous Section are
not lessened by this point.  On the contrary, removing it increases
significantly the goodness of the EBTF model with respect to the BTF
one.


\subsection{Independent sample}
\begin{figure}
  \resizebox{\hsize}{!}{\includegraphics[]{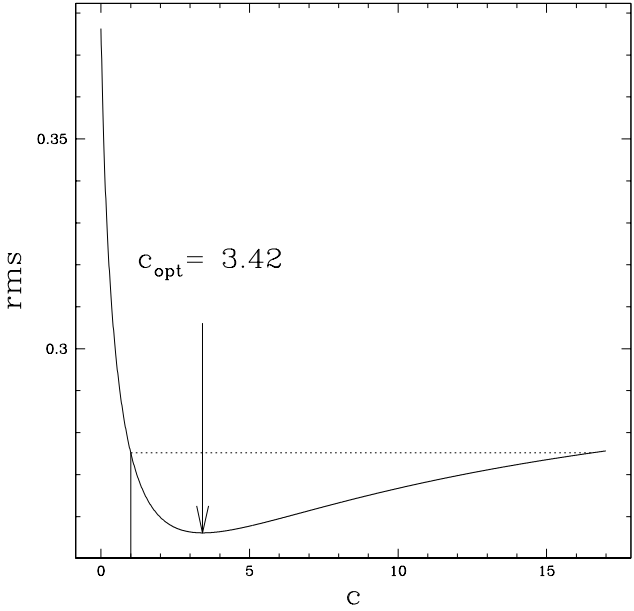}}
  \caption{Rms of the linear least-squares fit as a function of the 
    parameter $c$ (see Eq.~\ref{eq1}), for the sub-sample of \citet{sanders02}.
    The dotted line indicates the rms value for $c=1$.}
  \label{opt0_sanders}
\end{figure}

In order to check the dependence of the previous results on MSBD's
particular sample, we have repeated the analysis of
Sect.~\ref{comparing} on another independent sample.  We have chosen
the sample proposed by \citet{sanders02} that contains 76 galaxies,
where the luminosities (in either B, R, K or H bands) as well as the
HI mass are provided.  Among those galaxies, we have removed 29 of
them that are contained in MSBD's data. They correspond to the UMa
sample of \citet{verheijen01}.  The average mass-to-light ratio for
each band has been chosen according to MSBD.  For the R band, the
missing value has been interpolated and set to 1.2.  The rms of the
data fit of Eq.~\ref{eq1} as a function of $c$ is displayed in
Fig.~\ref{opt0_sanders}.  The found optimal value $c_{\rm{opt}}=3.42$
($\rm{rms}=0.256$) is slightly higher than for MSBD's sample. Beyond
this value, the rms grows slowly up to $c=16.5$ where it is still
better than for $c=1$ ($\rm{rms}=0.275$).
The results of the extra sum-of-squares $F$-Test, AICc and BIC tests
are displayed in Table~\ref{tests_sanders}. Those tests compare the
BTF fit with the EBTF one and with the quadratic fit of
Eq.~(\ref{quadratic}).
\begin{table}
  \begin{tabular}{l  l }
    \hline
    \hline
    BTF vs EBTF & BTF vs ``quadratic" \\
    \hline\\
    
    \multicolumn{2}{l}{$F$-Test} \\ \hline
    $F= 6.81$           &       $F=2.03$\\
    $P= 1.24$\,\%       &       $P=16$\,\%\\ \\
    
    \multicolumn{2}{l}{AICc} \\ \hline
    $\mathcal{A}^{\rm{c}}_1=-114.72$          & $\mathcal{A}^{\rm{c}}_1=-114.72$\\
    $\mathcal{A}^{\rm{c}}_2=-119.08$          & $\mathcal{A}^{\rm{c}}_2=-114.44$\\
    $\mathcal{P}_{2,1}/\mathcal{P}_{1,2}=8.8$ & $\mathcal{P}_{2,1}/\mathcal{P}_{1,2}=0.87$\\ \\

    \multicolumn{2}{l}{BIC} \\ \hline
    $\mathcal{B}_1=-109.72$                   & $\mathcal{B}_1=-109.72$\\
    $\mathcal{B}_2=-112.63$                   & $\mathcal{B}_2=-107.99$\\
    $\mathcal{P}_{2,1}/\mathcal{P}_{1,2}=4.2$ & $\mathcal{P}_{2,1}/\mathcal{P}_{1,2}=0.42$\\ \\
    \hline\hline
    
  \end{tabular}
  \caption[]{Statistical comparison of the BTF model with the EBTF one, 
    and with the quadratic model, for Sanders \& McGaugh sub-sample data.}
  \label{tests_sanders}
\end{table}
While the trend is slightly less pronounced in this new but smaller
sample, the statistical tests confirm that the EBTF model is better
than the BTF one ($4$ to $9$ times). The choice of Eq.~(\ref{eq1}) is
again supported by the fact that the quadratic model (with the same
number of parameters than the EBTF one) does not improve the fit.

\subsection{Variable $M_\star/L$}
A questionable assumption included in the stellar mass derivation in
MSBD's sample is that the stellar $M_\star/L$ ratio is taken as
constant, while in a sample of spirals it can typically vary by a
factor of several, more in B than in the NIR bands.  

However, along the spiral sequence the rotational velocity is known to
correlate with the stellar $M_\star/L$ ratio: small late type spirals
typically rotate slower and are bluer (so have a lower stellar
$M_\star/L$ ratio) than massive and fast rotating late spirals.  As an
illustration, in Broeils' sample (1992), the $M_\star/L_{\rm B}$
ratio of 23 bright galaxies spanning the spiral sequence can be
reasonably represented by a linear regression:
\begin{equation}
{M_\star \over L_{\rm B} } = 4.00 \log V_{\rm max} - 5.77 \, .
\end{equation}

Supposing now that instead of a constant $M_\star/L$ ratio, one would
include the above more accurate statistical knowledge in the EBTF.
The former constant $M_\star/L$ in MSBD data for each galaxy would be
replaced by a primed one of the form:
\begin{equation}
{ M_\star^\prime \over L } = 
   \left(1 + d \log {V_{\rm c} \over 125} \right) 
        { M_\star \over L } \, ,
\end{equation}
where the constant 125 comes from the interval middle value of the log of
velocities (in km/s) in MSBD's sample; $d$ is an additional parameter
that might also improve the BTF rms.  

The alternative EBTF fit now reads:
\begin{equation}
  \log 
    \left[ 
       \left( 1 + d \log { V_{\rm c} \over 125 } \right)  
       M_\star + c M_{\rm HI}
    \right] 
    \approx a + b \log V_{\rm{c}}\, , 
  \label{eq10}
\end{equation} 
where here $M_\star$ is meant not as the true stellar mass, but as the
quantity provided in MSBD's sample as the stellar mass estimate at
constant $M_\star/L$ ratio. 

Repeating the analysis of Sect.~3.1, we search for the best pair of
parameters $(c_{\rm opt}, d_{\rm opt})$ that minimizes the rms fit.
Interestingly, the best fit does indeed require a positive and
suitable $d$ with regard to the observational data, where $M_\star/L$ is
correlated with the rotational velocity.  This has the effect of
slightly increasing $c$.  However, the improvement in rms is very
modest.  The best pair is $c_{\rm opt} = 3.07$, $d_{\rm opt}= 0.48$,
for $\rm{rms}=0.3162$, instead of $c_{\rm opt} = 2.98$, ($d_{\rm
  opt}=0$) and $\rm{rms}=0.3164$ for EBTF.  The iso-values of
$\rm{rms}(c,d)$ around the minimum are strongly elongated, almost
straight, parallel to the $d$ axis, which already means that the most
relevant minimizing parameter is $c$.

With the tests of Sect.~3.2, we can now evaluate quantitatively
whether $d$ is a significant additional parameter.  For all three
tests the answer is clearly that $d$ does not bring any significant
improvement: for the $F$-test $P=0.61$, for AICc the information ratio
is only $\mathcal{P}_{2,1}/\mathcal{P}_{1,2} = 0.401$, and for BIC
even lower, $\mathcal{P}_{2,1}/\mathcal{P}_{1,2} =0.073$.

The results can be summarized concisely: the principal effect of a
variable $M_\star/L$ is to somewhat \textit{increase} $c_{\rm opt}$
from 2.98 to 3.07, while suggesting a fairly reasonable $M_\star/L$
variation over the spiral sequences. It especially shows that $c$
is a much more relevant parameter than $d$.  


\section{Discussion and conclusion}


An analysis of the baryonic Tully-Fisher relation of
MSBD's sample shows that a better correlation is obtained by
multiplying the total HI mass by a factor of around 3, suggesting that a
substantial amount of baryons would remain to be found in spirals.

The optimal factor around 3 is typically lower than the scaling factor
fitting well the total dynamical mass, $5-15$ \citep{hoekstra01}.
There is room for another component of a different nature, perhaps
non-baryonic and not located in the disk, still necessary to explain
the rest of the dynamical mass.  This point is supported by the recent
work of \citet{revaz04} showing that the frequent observations of
warps in spiral galaxies is the natural result of bending
instabilities, if, for a galaxy such as the Milky Way, the disk mass
within $30\,\rm{kpc}$ represents $\sim 60$\% of the total mass,
leaving $\sim 40$\% for a conventional non-baryonic dark halo.
However, one should keep in mind that the range of sub-optimal $c$
which still improve the original BTF relation rapidly widens toward
values up to about $11-16$, therefore including the full dynamical
mass may still improve the BTF relation.

In a first step the $M_\star/L$ ratio of the stellar component has
been supposed constant.  In a second step it has been allowed to vary
according to a more realistic trend where $M_\star/L$ is linear in
$\log V_{\rm c}$.  The statistical tests show that a variable
$M_\star/L$ is compatible with the observations and requires somewhat
\textit{more} dark baryons proportional to HI. More clearly the tests
demonstrate that a variable $M_\star/L$ ratio is a secondary factor with
regard to the mass proportional to the HI factor.

The reason why the Tully-Fisher relation should be a power law between
the disk baryonic mass and the disk rotational velocity remains
unclear, because the improvement found on the TF relation still does
not explain its physical origin; it just points toward an explanation
based on the disk baryonic content instead of either only the stellar
or only the detected baryonic content.  Dimensional arguments to
explain the TF relation by internal physical factors in self-regulated
gravitating disks, instead of initial conditions, were given in
\cite{pfenniger91} where it was pointed out that in pure
self-regulated disks the gravitational power $\propto v_{\rm rot}^5/G$
that a galaxy can exchange with stars appears comparable to the
stellar mechanical power, about $1/10$ of the stellar luminosity in
star forming disks.  Remarkably, this estimate of the disk internal
power provides a fair zero point of the TF relation.  The important
consequence is that the mechanical energy output coming mostly from
massive stars appears sufficient in magnitude to modify
secularly the global parameters of a disk galaxy.  Therefore to
understand spirals, their internal physics appears more important than the
initial conditions of formation.

This point is consistent with the unified picture of disk galaxies
presented in \cite{pfenniger04} where bars, spirals and warps result
from horizontal and vertical instabilities triggered by the constant
energy dissipation.  The basic properties of these systems may be
summarized by an interplay of cooling and self-gravitation.  In
dissipative self-regulated gravitating disks, more than the initial
conditions, the internal physics is crucial to understand their
structure and evolution.  The energy dissipated by radiative cooling,
on one hand, drives galaxies towards flat disk shapes, but, on the
other hand, also drives them toward gravitational instability
thresholds beyond which horizontal and vertical gravitational
instabilities are spontaneously triggered, heating the disk both in
the horizontal and vertical directions.  This purely dynamical
feed-back mechanism is supplemented by the heating caused by stellar
activity \citep{immeli04}, which results from the turbulence generated
in spiral arms by the larger-scale dynamical heating.  The galactic
disks appear then as tightly self-regulated structures that remain in
a marginally stable state as long as the cooling agent, the gas,
exists.  This seems to us a suitable starting point for finding a
better physical explanation of the TF relation.


\begin{acknowledgements}
  We thank Fr\'ed\'eric Pont and Raoul Behrend for useful discussions
  about statistical methods on model comparison.  We thank also the
  referee for critical but useful comments.  This work has been
  supported by the Swiss National Science Foundation.
\end{acknowledgements}


\begin{thebibliography}{}

\bibitem[Afonso et al.(2003)]{afonso03} 
Afonso, C., Albert, J.N., Andersen, J., et al.(EROS coll.)
2003, A\&A, 400, 951

\bibitem[Akaike(1974)]{akaike74}
Akaike, H. 
1974, IEEE Trans. Auto. Control, 19, 716

\bibitem[Alcock et al.(2001)]{alcock01}         
Alcock, C., Allsman, R.A., Alves, D.R., et al.(MACHO coll.)
2001, ApJ, 550, 169L

\bibitem[Bosma(1981a)]{bosma81a}        
Bosma, A. 
1981, AJ, 86, 1825

\bibitem[Boulanger(2004)]{boulanger04} 
Boulanger, F. 
2004, in ``Penetrating Bars through Masks of Cosmic Dust: The Hubble
Tuning Fork Strikes a New Note", D. Block et al. (eds.), Kluwer, in
press

\bibitem[Brockwell \& Davis(2002)]{brockwell02}  
Brockwell, P.J., Davis, R.A.                            
2002, Introduction to Time Series and Forecasting, Springer, Berlin

\bibitem[Broeils(1992)]{broeils92}      
Broeils, A., Dark and visible matter in spiral galaxies,
1992, PhD thesis, 
Rijksuniversiteit, Groningen 

\bibitem[Burnham \&  Anderson(2004)]{burnham04}
Burnham, K. P., Anderson, D.R. 
2004, Model selection and inference:
a practical information-theoretic approach. Springer-Verlag, New York

\bibitem[Combes(2004)]{combes04}                
Combes, F. 
2004, in : The Cold Universe,
Pfenniger, D., Revaz, Y. (eds.), Berlin,
Saas-Fee Advanced Course 32, Springer, Berlin, p. 105

\bibitem[de Paolis et al.(1995)]{depaolis95}  
De Paolis, F., Ingrosso, G., Jetzer, P., Roncadelli, M.
1995, A\&A, 296, 567

\bibitem[Fuchs(2003)]{fuchs03}
Fuchs, B. 
2003, Astroph. Space Sci. 284, 719

\bibitem[Gerhard \& Silk(1996)]{gerhard96}  
Gerhard, O., Silk, J.            
1996, ApJ, 472, 34      

\bibitem[Henriksen \& Widrow(1995)]{henriksen95}  
Henriksen, R.N., Widrow, L.M.            
1995, ApJ, 441, 70      

\bibitem[Hoekstra et al.(2001)]{hoekstra01}     
Hoekstra, H., van Albada, T.S., Sancisi, R. 
2001, MNRAS, 323, 453  
                                                
\bibitem[Immeli et al.(2004)]{immeli04} 
Immeli, A., Samland, M., Gerhard, O., Westera, P.
2004, A\&A, 413, 547

\bibitem[Kalberla(2003)]{kalberla03}                 
Kalberla, P.M.W. 
2003, \apj, 588, 805    
        
\bibitem[Liddle(2004)]{liddle04}
Liddle, A.R. 
2004, MNRAS, 351, L49
        
\bibitem[Lupton(1995)]{lupton95}  
Lupton, R.           
1993, Statistics in theory and practice, Princeton Univ. Press 

\bibitem[Masset \& Bureau(2003)]{masset03}      
Masset, F.D., Bureau, M.
2003, ApJ, 586, 152
   
\bibitem[Mayer \& Wadsley(2004)]{mayer04}
Mayer, L., Wadsley, J. 
2004, MNRAS, 347, 277
     
\bibitem[McGaugh et al.(2001)]{mcgaugh00}  
McGaugh, S.S., Schombert, J.M., Bothun, G.D., de Blok, W.J.G.             
2000, ApJ, 533, 99L   (MSBD)  

\bibitem[Pfenniger(1991)]{pfenniger91}      
Pfenniger, D. 
1991, in Dynamics of Disc Galaxies, B. Sundelius (ed.), G\"oteborg,
389
 
\bibitem[Pfenniger \& Combes(1994)]{pfenniger94b}      
Pfenniger, D., Combes, F.
1994, A\&A, 295, 94

\bibitem[Pfenniger, Combes \& Martinet(1994)]{pfenniger94a}    
Pfenniger, D., Combes, F., Martinet, L.
1994, A\&A, 285, 79

\bibitem[Pfenniger \& Revaz(2004)]{pfenniger04} 
Pfenniger, D., Revaz, Y.  
2004, in ``Penetrating Bars through Masks of Cosmic Dust: The
Hubble Tuning Fork Strikes a New Note", D. Block et al. (eds.),
Kluwer, in press (astro-ph/0406578)

\bibitem[Revaz \& Pfenniger(2004)]{revaz04}   
Revaz, Y., Pfenniger, D. 
2004, A\&A, in press (astro-ph/0406339)

\bibitem[Sanders \& McGaugh(2002)]{sanders02}
Sanders, R.H., McGaugh, S.S.
2002, ARA\&A, 40, 263

\bibitem[Sakamoto, Ishiguro \& Kitagawa(1986)]{sakamoto86}
Sakamoto, Y., Ishiguro, M., and Kitagawa, G. 
1986, Akaike Information Criterion Statistics, Reidel
     
\bibitem[Schwarz(1978)]{schwarz78}
Schwarz, G. 
1976, Annals of Statistics, 5, 461
     
\bibitem[Tully \& Fisher(1977)]{tully77}
Tully, B., Fisher, J. R.  
1977, A\&A 54, 66 (TF)

\bibitem[Verheijen(2001)]{verheijen01}
Verheijen, M.A.W.
2001, ApJ, 563, 694

\end{thebibliography}
\end{document}